\documentclass[pra,twocolumn,showpacs,preprintnumbers,amsmath,amssymb,floatfix]{revtex4}
\usepackage{epsfig}
\usepackage{graphicx}
\usepackage{amssymb}
\usepackage{color}

\begin{document}

\title{Speeding up critical system dynamics through optimized evolution.}

\author{Tommaso Caneva$^{1,2}$}
\author{Tommaso Calarco$^2$}
\author{Rosario Fazio$^{3}$}
\author{Giuseppe E. Santoro$^{1,4,5}$}
\author{Simone Montangero$^{2}$}

\affiliation{$^1$International School for Advanced Studies (SISSA), Via Beirut 2-4,
  I-34014 Trieste, Italy\\
$^2$Institut f\"ur Quanteninformationsverarbeitung,
  Universit\"at Ulm, D-89069 Ulm, Germany\\
$^3$NEST, Scuola Normale Superiore $\&$ Istituto di Nanoscienze - CNR, 
  Piazza dei Cavalieri 7, I-56126 Pisa, Italy\\
$^4$CNR-INFM Democritos National Simulation Center, 
  Via Beirut 2-4, I-34014 Trieste, Italy\\
$^5$International Centre for Theoretical Physics (ICTP), 
  P.O.Box 586, I-34014 Trieste, Italy}

\date{\today}

\begin{abstract}
The number of defects which are generated on crossing a quantum phase transition can be minimized by 
choosing properly designed time-dependent pulses. In this work we determine what are the ultimate
 limits of this  optimization. We discuss under which conditions the production of defects across the 
phase transition is vanishing small. Furthermore we show that the minimum time required to enter this 
regime is $T\sim\pi/\Delta$, where $\Delta$ is the minimum spectral gap, unveiling an intimate connection 
between an optimized unitary dynamics and the 
intrinsic measure of the Hilbert space for pure states. Surprisingly, the dynamics is non-adiabatic,
this result can be understood by assuming a simple two-level dynamics for the many-body system. Finally
we classify the possible dynamical regimes in terms of the action
$s=T\Delta$.
\end{abstract}

\maketitle

\begin{figure}
\epsfig{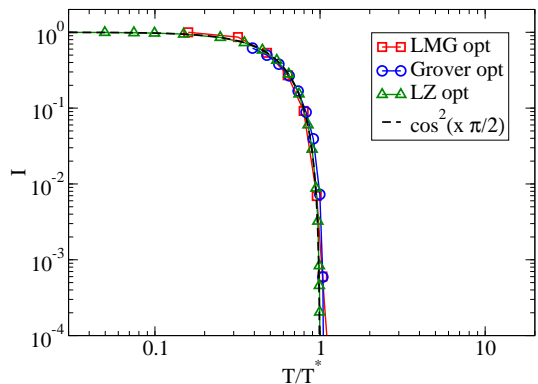}
\caption{(Color online) Infidelity $I$ as a function of the adimensional scaling variable $T/T^*$
for the LMG (red squares), Grover (blue circles) and the LZ model (green triangles). 
Data correspond to half of the maximum size analyzed ($N=64$).}
\label{L64_comparison_fig}
\end{figure}

%
The rapid progress in the experimental realization and manipulation of quantum 
systems~\cite{Bloch_RMP08} 
is opening the rich and intriguing perspective of the exploitation of 
quantum physics to realize quantum technologies 
like quantum simulators~\cite{Feynman_IJTP82} and quantum
computers~\cite{Lloyd_NAT00,Farhi_SCI01}. These achievements
pave the way to the simulation of
condensed matter systems giving the possibility of
studying different states of matter in controlled experiments~\cite{Baumann_NAT10}. 
Despite the impressive results obtained so far, this is a formidable
technological and theoretical challenge due to the complexity of the
systems in analysis and the experimental requirements. 
Indeed, the level of control needed on the quantum system is unprecedented: one
should be able to prepare a system in a desired initial state, perform
the desired evolution and finally measure the state in a very precise way. 
Moreover, the whole experiment should be performed faster than the system
decoherence time that eventually will destroy any quantum information
capability.\\
Quantum optimal control (OC) theory,
the study of optimization  strategies to improve 
the outcome of a quantum process, can be 
an extremely powerful tool to cope with these
issues~\cite{Krotov:book,Sola_JPCA98,Khaneja_JMR05,Montangero_PRL07,Brif_NJP10}. 
It allows not only to optimize the desired experiment outcome but also
to speed up the process itself. Traditionally employed in atomic and molecular
physics~\cite{Peirce_PRA88,Calarco_PRA04}, OC
has been recently applied with success to the optimization of the
dynamics of many-body systems~\cite{Caneva_PRL09,Doria_10:preprint,Rahmani_10:preprint}, 
allowing to achieve the ultimate bound imposed by quantum mechanics,
the so called \emph{quantum speed limit} (QSL)~\cite{Giovannetti_PRA03}.
Indeed as intuitively suggested by the time-energy uncertainty 
principle, the time required by a state to reach another distinguishable state has to be 
longer than the inverse of its energy fluctuations~\cite{Aharonov_PR61}. 
This implies that a quantum system cannot evolve at an arbitrary speed
in its Hilbert space, but a minimum time is required to perform a 
transformation between orthogonal 
states~\cite{Battacharyya_JPA83,Margolus_PD98,Levitin_PRL09,Smerzi_10:preprint,Jones_PRA10}.
\begin{figure}
\epsfig{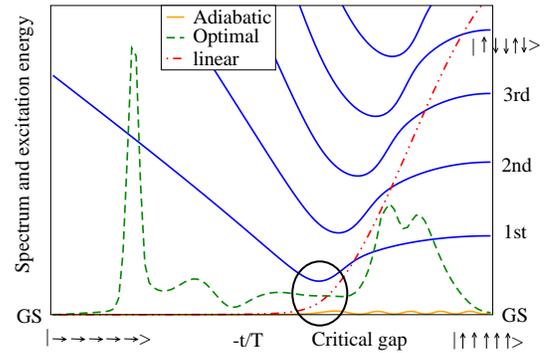}
\caption{(Color online) Instantaneous excitation energy in the LMG model for an optimized (green dashed line, total time 
$T\sim T_{\rm QSL}$), a non 
optimized (red dot-dashed line, 
$T\sim T_{\rm QSL}$) and a linear adiabatic process (orange continuous line, 
$T\gg T_{\rm QSL}$). 
Continuous (blue) lines represents the lowest energy levels as a function of the driving field
$\Gamma=-t/T$.}
\label{cartoon_fig}
\end{figure}
For time-independent Hamiltonians this bound
has been exactly determined~\cite{Giovannetti_PRA03};
the QSL has been formally generalized
also to time-dependent Hamiltonians, but so far has been computed only
in a few simple cases~\cite{Anandan_PRL90,Pfeifer_PRL93,Carlini_PRL06,Rezakhani_PRL09,Caneva_PRL09}.\\
A still unexplored, although relevant question is how the dynamical crossing of a quantum
phase transition (QPT) affects this fundamental bound.
QPTs indeed represent a dramatic change in the low energy sector of a quantum system and their presence
strongly influence its dynamics. The study of the dynamical crossing of phase transitions was initially 
considered in cosmology to investigate the formation of the universe\cite{Zurek_NAT85}.
More recently with the developement of the quantum annealing\cite{Brooke_SCI99} and adiabatic quantum 
computation\cite{Farhi_SCI01}, a renewed interest has been devoted to the 
subject in condensed matter and quantum 
information\cite{Zurek_PRL05,Polkovnikov_NAT08,Dziarmaga_AP10,Polkovnikov_10:preprint}.
Here we investigate for the first time the QSL of the dynamics of a first order 
QPT in the adiabatic version of Grover's search algorithm (GSA)~\cite{Roland_PRA02,Rezakhani_PRA10} and 
of a second order 
QPT~\cite{Sachdev:book} in the Lipkin-Meshkov-Glick (LMG) model. Specifically we consider the 
problem of converting the ground state on one side of the critical point  
into the ground state on the opposite side in the fastest
and most accurate way by selecting an optimal time-dependence
of the control field. We emphasize 
here that the evolution induced by the optimized field is non-adiabatic, as shown in 
Fig.~\ref{cartoon_fig}, where the scenario is reproduced for the LMG model, and an adiabatic 
and an optimal evolution are compared: an adiabatic strategy (orange continuous line) 
turns out to be effective only for a very large total evolution time $T$ (namely 
$T\gg\hbar\Delta^{-1}$, with $\Delta$ being the minimum spectral gap, as required by the adiabatic 
theorem\cite{Messiah:book}). When the total time is reduced, as realistically required in experiments
in order to preserve the phase coherence, adiabaticity fails, leading to an excited
state far from the target (red dot-dashed line). However relaxing the constraint of 
adiabaticity and allowing a more general non adiabatic evolution, with OC it is possible to reach 
the desired goal with a fast dynamics (green dashed line).
Quite surprisingly our study shows that the outcome of the 
dynamical process optimization for the many-body systems analyzed is 
independent from the specific model and
analogous to that of a two-level system, as sketched through the 
good rescaling of the data in 
Fig.~\ref{L64_comparison_fig}. We interpret this result 
as the natural manifestation of the intrinsic metric of the Hilbert space for pure 
states~\cite{Wootters_PRD81,Anandan_PRL90}, as discussed in Sec.~\ref{metric:sec}.
Furthermore, studying the QSL as a function 
of the system size, we show that the speed up 
obtained by the adiabatic GSA~\cite{Grover_PRL97,Roland_PRA02} 
can be reproduced and extended to other models with optimized, non-adiabatic evolutions.
Finally, we introduce the action $s=T\Delta$ as a parameter to characterize
the evolution of a quantum system and we find that the QSL  
identifies a new dynamical regime, as discussed in Sec.~\ref{classification:sec} and
summarized in Fig.~\ref{comparison}.

\section{Models and optimization}

%
\begin{table*}
\begin{tabular}{|c|c|c|c|c|c|c|}
\hline
Model & $H$ & $|\psi _i\rangle$ & $|\psi _G\rangle$ & $\Delta$ & $s^*_{Lin}$ & $s^*_{Opt}$ \\ \hline
GSA & $(1-\Gamma (t))(\mathbf{1}-|\psi_i\rangle \langle|\psi _i|)+\Gamma (t)(\mathbf{1}-|\psi_G\rangle \langle|\psi _G|)$ & $\sum _i ^N|i\rangle/\sqrt{N} $ & $|10...0\rangle$ & $N^{-1/2}$ & $f_1(I) N^{1/2}$ & $\stackrel{N\gg 1}{\longrightarrow}\pi$ \\\hline
LMG & $-(N^{-1})\sum ^N _{i< j}\sigma_i ^x \sigma_j ^x -\Gamma (t) \sum _{i}^N \sigma_i ^z$ & $|\uparrow ...\uparrow\rangle _z$ & $|\leftarrow ...\leftarrow\rangle _x,|\rightarrow ...\rightarrow\rangle _x$  & $N^{-1/3}$ & $f_2(I)N^{1/3}$ & $\pi$ \\\hline
LZ & $\Gamma(t) \sigma ^z + \omega \sigma ^x$ & $|\uparrow\rangle _z$ & $|\downarrow\rangle _z$ &$\Delta$ &$(-4\ln (I)/\pi)\Delta ^{-1}$ & $\pi$ \\
\hline
\end{tabular}
\caption{Hamiltonian models and predicted scalings 
  in the LZ approximation for the linear and optimal quenches. They
  should be compared with the results of
  Fig.~\ref{linear_scale_pic_fig}. The functions $f_\imath(I)$
  diverge for the infidelity $I\to 0$.}
\label{scalings}
\end{table*}
%
We study two paradigmatic critical systems, the adiabatic GSA~\cite{Roland_PRA02}
and the LMG model~\cite{Botet_PRB83}
and we compare them with the Landau-Zener (LZ) 
model to better understand the physics of the process. 
The GSA Hamiltonian is given by:
\begin{eqnarray}
H^{\rm GSA}&=&(1-\Gamma (t))(I-|\psi_i\rangle \langle|\psi _i|)\nonumber\\
&+&\Gamma (t)(I-|\psi_G\rangle \langle|\psi _G|)
\end{eqnarray}
where the initial state is an equal superposition of all $N$ basis states $|i\rangle$, i.e. 
$|\psi_i\rangle =(\sum_i ^N|i\rangle)/\sqrt{N}$
and the final target is the specific marked state we want to exctract from the database
(in our simulations $|\psi_G\rangle =|10...0\rangle$ without loss of generality).
The system undergo a 1st order QPT
at a critical value of the transverse field $\Gamma_ c =0.5$ (from
now on we set $\hbar=1$). The gap between the ground state and first
excited state closes polynomially with the size at the critical point: $\Delta _{\rm GSA}\sim
N^{-1/2}$.\\
The LMG Hamiltonian instead is written as:
\begin{eqnarray}
H^{\rm LMG}=-\sum ^N _{i< j}J_{ij}\sigma_i ^x \sigma_j ^x -
    \Gamma (t) \sum _{i}^N \sigma_i ^z ,
\label{ham:eq}
\end{eqnarray}
where $N$ is the number of spins, $\sigma _i^{\alpha}$'s ($\alpha =x,y,z$ ) are the Pauli matrices on 
the $i$th site and $J_{ij}=1/N$ (infinite range interaction). The system undergo a 2nd order QPT
from a quantum paramagnet to a quantum ferromagnet
at a critical value of the transverse field $|\Gamma_ c| =1$. The gap between the ground state and first
excited state closes polynomially with the size at the critical point: $\Delta _{\rm LMG}\sim
N^{-1/3}$. We chose as initial state the ground state (GS) at $\Gamma _i\gg 1$,
i.e. the state in which all the spins
are polarized along the positive $z$-axis (paramagnetic phase).
As target state has been chosen the GS at $\Gamma=0$.\\
Finally the LZ
Hamiltonian that we use as a reference model is 
\begin{eqnarray}
H^{\rm LZ}= \Gamma(t) \sigma _z + \omega \sigma _x,
\end{eqnarray}
where the off-diagonal terms
give the amplitude of the minimum gap $\Delta _{\rm LZ} = 2 \omega$ at the anticrossing point $\Gamma=0$, here
assumed to be at
$t=0$~\cite{Zener_PRS32,Zurek_PRL05}. 
In this case the initial state is the GS for
$\Gamma(-T/2)= -\Gamma_0$ and the target is the GS for
$\Gamma(T/2)=\Gamma_0$, that is ---in this effective model--- to transform the initial GS into the
initial excited state in an optimal and fastest way.
The systems analysed have been summarized in the left side of Table~\ref{scalings}.\\
For all the models considered
our goal is to find the optimal driving control field $\Gamma(t)$ to
transform the initial in the goal state in a given total time $T$. 
At the limit when the gap closes (the thermodynamical limit for GSA and LMG)
adiabatic dynamics is forbidden in finite time due to the adiabatic condition $T\gg
\Delta^{-1}$~\cite{Messiah:book}: however, for finite size systems,
an adiabatic strategy might be successful.
Here we relax the adiabaticity condition, exploring a different  
regime of fast non adiabatic transformations. Given the 
total evolution time $T$, we use optimal control through the Krotov's
algorithm to find the optimal control field $\Gamma (t)$ 
to minimize the infidelity $I(T)=1-|\langle \psi _G|\psi
(T)\rangle|^2$ at the end of the evolution, i.e. the
discrepancy between the final and the goal state~\cite{Krotov:book}. 
The determination of $\Gamma _{\rm opt}(t)$ can be recast in a minimization
problem subject to constraints determined by looking for the
stationary points of a functional $\mathcal{L}[\psi,\dot{\psi},\chi,\Gamma]$
in which the auxiliary states $|\chi (T)\rangle = |\psi _G\rangle
\langle\psi _G|\psi(T)\rangle $ play the 
role of a continuous set of Lagrange 
multipliers to impose the fulfillment of the Schr\"odinger equation at
each time during the dynamics, as described in details
in~\cite{Krotov:book,Calarco_PRA04,Montangero_PRL07}.

\section{Results}
\subsection{Hilbert space metric and optimization}
\label{metric:sec}
Previous studies~\cite{Caneva_PRL09} revealed that only when the 
total evolution time exceeds a certain threshold, by iterating 
the algorithm it is possible to reduce arbitrarily 
the value of the final infidelity $I$. 
%
\begin{figure}
\epsfig{file=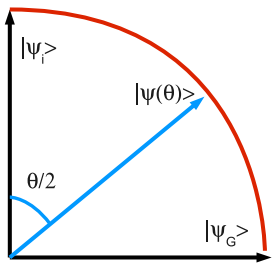,width=5cm,angle=0,clip=}
\caption{(Color online) Schematic representation of the ray $|\psi _{\theta}\rangle$ evolving
along the geodesic connecting the initial state $|\psi _i\rangle$ and the target $|\psi _G\rangle$.}
\label{metric_fig}
\end{figure}
\begin{figure}
\epsfig{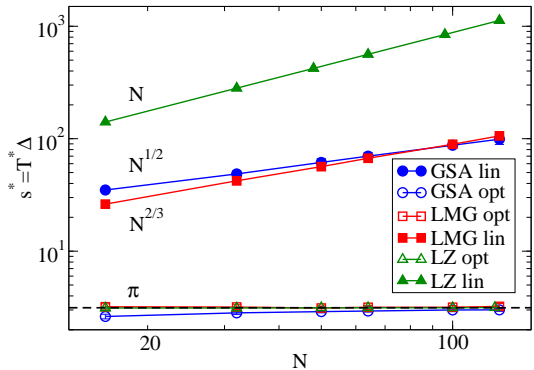}
\caption{(Color online) The action $s^*=T^*\Delta$ 
as a function of the size $N$ (for the LZ model we define an effective size $N=\Delta ^{-1}$), 
where $\Delta$ is the minimum spectral gap and 
$T^*$ the time required to reach an infidelity $I^*\sim 10^{-3}$
for the linear (full symbols) and optimized (empty
symbols) driving field for the LMG (red squares), 
Grover (blue circles) and the LZ model (green triangles). 
}
\label{linear_scale_pic_fig}
\end{figure}
In order to identify such a threshold, we fix
a target value of the infidelity $I^*\sim 10^{-3}$ and we determined the 
minimum total evolution time $T^*$ for which it is possible to satisfy our goal. 
In Fig.~\ref{L64_comparison_fig} we show the value of the infidelity 
for the optimized process as a function of the rescaled time $T/T^*$ for
different models. 
The first observation is the presence of a sharp threshold,
thus, $T^*$ can be considered a reliable estimate of the 
QSL for the process considered.
The second striking feature is the rescaling of the data onto the
function $I=\cos ^2(T/T^*)$.
We interpret this general behavior as a manifestation of 
the Fubini-Study metric~\cite{Wootters_PRD81,Anandan_PRL90} in the Hilbert space. 
The presence of such a metric for pure states can be demonstrated
following two independent approaches: Wootters in Ref.~[\onlinecite{Wootters_PRD81}]
obtained it just from statistical considerations; Anandan and Aharonov instead in 
Ref.~[\onlinecite{Anandan_PRL90}] derived it generalizing the concept
of the geometrical Berry's phase to generic non adiabatic evolutions.
According to the Fubini-Study metric the distance separating two arbitrary pure states is 
given by the angle between the corresponding rays,
$\theta _G/2=\arccos |\langle\psi _i|\psi_G\rangle|$. For orthogonal states this distance 
is maximal and given by $\theta _G=\pi$.
The shortest path connecting the states $|\psi_i\rangle$ and $|\psi_G\rangle$
is then represented by a geodesic in the ray subspace.
We can indicate a ray evolving along such a geodesic with $|\psi_{\theta}\rangle$, 
where $0<\theta< \theta _G$, so that $|\psi _0\rangle=|\psi_i\rangle$ and 
$|\psi _{\theta _G}\rangle=|\psi_G\rangle$, as sketched in Fig.~\ref{metric_fig}.
For such a ray it turns out 
$|\langle\psi _{\theta}|\psi_G\rangle|^2 =\cos ^2((\theta_G-\theta)/2)$
or, for the infidelity, $I_{\theta}=\sin ^2((\theta _G-\theta)/2)$. 
Numerically can be verified that for states on opposite sides of
a QPT $\theta _G\simeq\pi$; substituting this value in the expression
of the infidelity,  
we obtain $I_{\theta}\simeq \cos ^2(\theta/2)$. Making the identification $\theta=\pi T/T^*$
this last formula is in perfect 
agreement with the data of 
Fig.~\ref{L64_comparison_fig} and with the results for a two-level
system in Ref.~[\onlinecite{Caneva_PRL09}]. An optimized evolution then
can be interpreted as a uniform motion along a geodesic with speed
$\pi/T^*$.
%
\subsection{Dynamical regime classification}
\label{classification:sec}
In order to establish a classification of dynamical regimes and to
understand the speed up that optimized evolutions 
gain with respect to non optimized pulses, we introduce  
the action $s=T\Delta$, obtained through the 
product of the total evolution time with the minimum spectral gap.
Notice this is a quite natural way to characterize a dynamical process: 
$s\gg 1$ corresponds to a slow evolution, for which in principle 
an adiabatic dynamics could be achieved; $s\leq 1$ instead characterizes 
a fast evolution, for which adiabaticity is strictly forbidden.
In Fig.~\ref{linear_scale_pic_fig} the data for optimal driving
fields (empty symbols) are compared with data obtained with a linear time
dependence (full symbols) $\Gamma (t)=t/T$ for the GSA, the LMG 
and the LZ models.
We report the product $s^*=T^* \Delta$
as a function of the size $N$, 
being $\Delta$ the minimum spectral gap and $T^*$ the minimal time required to reach an
infidelity $I^*\sim 10^{-3}$. 
As can be clearly seen in Fig.~\ref{linear_scale_pic_fig}, 
a linear time-dependent $\Gamma (t)$ results in an
action $s^*$ increasing with the system size, implying  that
$T^*\sim \Delta ^{-\alpha}$ with $\alpha > 1$ (full symbols). On the contrary, 
the action $s^*$ remains below the value $\pi$ after the
optimization 
(empty symbols). 
Notice that the optimal action is reduced by one to two orders
of magnitude in Fig.~\ref{linear_scale_pic_fig}, but in general $s^*_{Opt} \sim \pi$ 
and $s^*_{Lin}\rightarrow\infty$ for $I^*\rightarrow 0$. A simple interpretation
of the scalings reported in the picture, can be understood as follows.
For a linear quench, in a first approximation, we can assume that the main
contribution to the infidelity comes only from the first excited state~\cite{Zurek_PRL05},
and the Landau-Zener formula~\cite{Zener_PRS32} can be used to give an estimate of the excitation
probability, i.e. the infidelity: 
$\mathcal{I}=\exp (-\beta \Delta^2 T)$, with $\beta= {\rm const}$, 
so that by fixing an arbitrary (but small) value for the infidelity 
$\mathcal{I}^*$, we have $T^*\sim \Delta^{-2}$ or $s^*\sim \Delta^{-1}$.
By inserting the gap dependence on the size,
the scalings reported 
in Table~\ref{scalings} are obtained for the models considered in this work:
they are in almost perfect agreement with the 
numerical data reported in Fig.~\ref{linear_scale_pic_fig}, where
$s^*_{Lin}$ is increasing with the size.
The only discrepancy is the scaling of the linear LMG
model due to the fact that the simple linear LZ approximation fails. 
Indeed, here we have that $\beta=\beta(N) \sim
N^{-1/3}$, resulting in $T^* \sim \Delta^{-3}$, that is, $s_{Lin}^*= N^{2/3}$.
For the optimized process instead, the optimal value 
$s^*_{Opt}= \pi$ corresponds to a Rabi oscillation between the initial and the target state
at a frequency $\omega _R=\Delta/2$, as clearly shown in Fig.~\ref{L64_comparison_fig}; 
or in other words, according to our geometrical interpretation, the optimal 
evolution can be seen as a motion along a geodesic connecting the initial and the 
target state at a constant speed proportional to $\Delta$~\cite{Carlini_PRL06}; an intuitive
explanation is provided in the Appendix.
As shown in Fig.~\ref{linear_scale_pic_fig}, 
the speed-up we have obtained in our analysis
is analogous to the speed-up reached through the Grover's 
quantum adiabatic 
algorithm~\cite{Roland_PRA02}, from a quadratic to a linear dependence 
on gap of the evolution time $T^*$. In the case of the LMG model 
the gain is stronger, from a cubic to a linear dependence,
outlining the fact that the limit of the optimization is set by
a constant value of the action $s^*=T^*\Delta$.
As a last remark, from the previous discussion it can be argued that
optimized evolutions achieve a substantial speed up only when the minimum 
gap closes polynomially with the size; in the case of an exponential  
closure, even an optimized process leads to a 
total evolution time exponentially diverging with 
$N$~\cite{Dziarmaga_PRB06,Caneva_PRB07,Amin_PRA09,Altshuler_09:preprint,Young_PRL10}.

\begin{figure}[t]
\epsfig{file=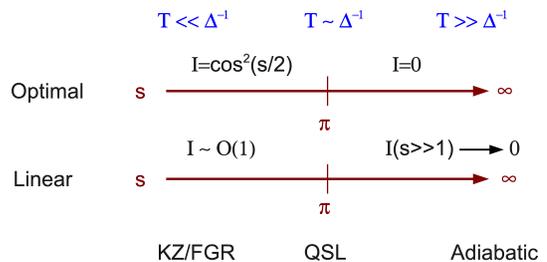,width=7cm,angle=0,clip=}
\caption{(Color online) The different dynamical regimes as a function of the
  action $s$.}
\label{comparison}
\end{figure}

We summarize the possible regimes of a quantum evolution in Fig.~\ref{comparison},
where the final infidelity and the timescale as a function of the action $s$
are shown. An optimized process is characterized by $I= \cos ^2(s/2)$ for $s\leq\pi$ and $I=0$ for $s>\pi$.
For a linear non-optimized pulse, $s \leq \pi$ indicates a fast evolution with high
defect production ($I\sim\mathcal{O}(1)$) accurately estimated by Kibble-Zurek (KZ) theory~\cite{Zurek_PRL05} 
or Fermi golden rule (FGR) approximation~\cite{Polkovnikov_NAT08}. On the contrary,
for $s\gg \pi$ adiabaticity can be achieved and the infidelity is asymptotically vanishing.
The QSL then clearly identifies the threshold between fast and adiabatic evolutions.

\section{Conclusion}
%
%
%
In conclusion, we estimated the time required by an optimized, non-adiabatic 
process to drive the ground state of a many-body system 
across a quantum phase transition. The behavior of the systems analyzed revealed
to be surprisingly similar to that one of a simple 2-level system; we explained
the phenomenon through the connection with the intrinsic geometry of the 
Hilbert space for pure state, 
interpreting the optimized process as an uniform motion along the geodesic connecting
the initial and target states, as summarized in Fig.~\ref{L64_comparison_fig}.
This result is of particular relevance because establishes a direct link between
quantum speed limit and Fisher information~\cite{Jones_PRA10} in the general setup of an evolution 
driven by a time dependent Hamiltonian. 
We demonstrated that the QSL for the dynamical processes analyzed scales as the 
inverse of the critical gap, significantly improving
the result obtained with a non optimized evolution. Such a speed-up is a general 
feature and can be interpreted as the extension of the Grover's algorithm speed-up to the 
models considered. 
Finally, introducing the action $s$, we provided a classification
of the possible QPT dynamics.  
We mention that the understanding of these
optimal process and of the fundamental timescales might be used to
develop new and more efficient optimization strategies also in quantum state preparation.

\emph{Acknowledgments}.--- 
%
We thank L. Viola and V. Giovannetti for useful discussions. We acknowledge financial
support by AQUTE, SFB TRR21, PICC, SOLID and BW-grid for computational resources.

\section*{Appendix}
%
\begin{figure}
\epsfig{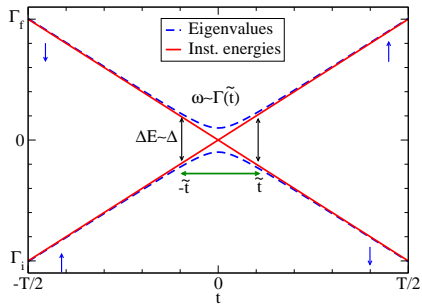}
\caption{(Color online) Instantaneous eigenvalues (blue dashed lines) 
and uncoupled energies of the spin-up, spin-down configurations
(red continuous lines) of the LZ model for  a linear quench, 
$\Gamma\sim t/T$. The time $\tilde{t}$ represents 
the boundary between Region I with high transformation rate 
($|t|\leq \tilde{t}$) and Region II where the system is in the
instantaneous ground state ($|t|\geq \tilde{t}$).}
\label{QSL_persp_LZ_lin_fig}
\end{figure}
In this Appendix we provide an intuitive explanation for the scaling
obtained in Sec.~\ref{classification:sec}.
The numerical results suggest that the optimized process is equivalent
to perform a rotation of the initial state into the target at 
a constant angular speed $\omega=\Delta/2$, where $\Delta=f(N)$
is the critical gap of the finite size many-body system in analysis.
In a first approximation we can recast the full many-body problem 
into a 2-level effective model, described with a LZ like 
Hamiltonian $H^{\rm LZ}[\Gamma (t)]$, where the spin-up and spin-down
states play respectively the role of the full many-body initial and target
state and the tunability of the diagonal element $\Gamma (t)$ mimics
the possibility of selecting the instantanous rotational axis.
Considering Fig.(\ref{QSL_persp_LZ_lin_fig}),
intuitively one can expect that the initial state does not change
significantly while its energy is much larger than the off-diagonal 
matrix elements, while viceversa, an efficient population transfer 
occurs when $\Gamma (t)\lesssim \omega$. Therefore the total
evolution can be approximated in two distinct regimes:
a first one for $|\Gamma(t)|\gg\omega$ in which $H^{\rm LZ}\sim
\Gamma(t)\sigma _z$  and the initial
state corresponds to the instantaneous GS. The
second regime, for $|\Gamma(t)|\ll\omega$ where the two levels are
highly coupled and $H^{\rm LZ}\sim \omega\sigma _x$.
The time $\tilde{t}$ marking the boundary between the two regimes is implicitely 
determined by the condition 
\begin{equation}
\Gamma (\tilde{t})\sim \omega,
\label{condition}
\end{equation}
as shown in Fig.(\ref{QSL_persp_LZ_lin_fig}). 
Under this approximation, the optimization problem is now easily solved.
It is indeed known that 
the fastest possible transition between two orthogonal states with
a fixed overlap $\omega$ is obtained through a Rabi oscillation, i.e. by applying 
$H=\omega\sigma _x$ for a time $T_c=\pi/2\omega$~\cite{Anandan_PRL90,Lloyd_NAT00}.
The complete rotation is then possible if the condition $T_c\sim \tilde{t}$
holds, where $\tilde{t}$ is obtained solving Eq.(\ref{condition}).
A linear time-dependence of the field, $\Gamma (t)=t/T$, 
gives $\tilde{t}\sim \omega T $, that is $T \sim \omega^{-2} \sim
\Delta^{-2}$, implying $s^*_{Lin} \sim \Delta^{-1}$.
The optimization of the time dependence of
$\Gamma (t)$ corresponds to the extension of the region of high 
transition rate, then increasing $\tilde{t}$: the best possible result is
clearly $\tilde{t}\sim T$, that is, the transformation is effective during the whole 
evolution. Finally setting this condition, we obtain 
$T \sim \pi/2\omega\sim \pi\Delta^{-1}$ and $s^*_{Opt} \sim \pi$.

\bibliographystyle{apsrev}

\end{document}